# A Personalized System for Conversational Recommendations

**Cynthia A. Thompson**                                    CINDI@CS.UTAH.EDU
*School of Computing*
*University of Utah*
*50 Central Campus Drive, Rm. 3190*
*Salt Lake City, UT 84112 USA*

**Mehmet H. Göker**                                    MGOKER@KAIDARA.COM
*Kaidara Software Inc.*
*330 Distel Circle, Suite 150*
*Los Altos, CA 94022 USA*

**Pat Langley**                                    LANGLEY@ISLE.ORG
*Institute for the Study of Learning and Expertise*
*2164 Staunton Court*
*Palo Alto, CA 94306 USA*

## Abstract

Searching for and making decisions about information is becoming increasingly difficult as the amount of information and number of choices increases. Recommendation systems help users find items of interest of a particular type, such as movies or restaurants, but are still somewhat awkward to use. Our solution is to take advantage of the complementary strengths of personalized recommendation systems and dialogue systems, creating personalized aides. We present a system – the ADAPTIVE PLACE ADVISOR – that treats item selection as an interactive, conversational process, with the program inquiring about item attributes and the user responding. Individual, long-term user preferences are unobtrusively obtained in the course of normal recommendation dialogues and used to direct future conversations with the same user. We present a novel user model that influences both item search and the questions asked during a conversation. We demonstrate the effectiveness of our system in significantly reducing the time and number of interactions required to find a satisfactory item, as compared to a control group of users interacting with a non-adaptive version of the system.

## 1. Introduction and Motivation

Recommendation systems help users find and select items (e.g., books, movies, restaurants) from the huge number available on the web or in other electronic information sources (Burke, 1999; Resnick & Varian, 1997; Burke, Hammond, & Young, 1996). Given a large set of items and a description of the user's needs, they present to the user a small set of the items that are well suited to the description. Recent work in recommendation systems includes intelligent aides for filtering and choosing web sites (Eliassi-Rad & Shavlik, 2001), news stories (Ardissono, Goy, Console, & Torre, 2001), TV listings (Cotter & Smyth, 2000), and other information.

The users of such systems often have diverse, conflicting needs. Differences in personal preferences, social and educational backgrounds, and private or professional interests are pervasive. As a result, it seems desirable to have *personalized* intelligent systems that process, filter, and display available information in a manner that suits each individual using them. The need for personalization has led to the development of systems that adapt themselves by changing their behavior based on the inferred characteristics of the user interacting with them (Ardissono & Goy, 2000; Ferrario, Waters, & Smyth, 2000; Fiechter & Rogers, 2000; Langley, 1999; Rich, 1979).

The ability of computers to converse with users in natural language would arguably increase their usefulness and flexibility even further. Research in practical dialogue systems, while still in its infancy, has matured tremendously in recent years (Allen, Byron, Dzikovska, Ferguson, Galescu, & Stent, 2001; Dybkjær, Hasida, & Traum, 2000; Maier, Mast, & Luperfoy, 1996). Today's dialogue systems typically focus on helping users complete a specific task, such as planning, information search, event management, or diagnosis.

In this paper, we describe a personalized conversational recommendation system designed to help users choose an item from a large set all of the same basic type. Our goal is to support conversations that become more efficient for individual users over time. Our system, the ADAPTIVE PLACE ADVISOR, aims to help users select a destination (in this case, restaurants) that meets their preferences.

The ADAPTIVE PLACE ADVISOR makes three novel contributions. To our knowledge, this is the first personalized spoken dialogue system for recommendation, and one of the only conversational natural language interfaces that includes a personalized, long-term user model. Second, it introduces a novel model for acquiring, utilizing, and representing user models. Third, it is used to demonstrate a reduction in the number of system-user interactions and the conversation time needed to find a satisfactory item.

The combination of dialogue systems with personalized recommendation addresses weaknesses of both approaches. Most dialogue systems react similarly for each user interacting with them, and do not store information gained in one conversation for use in the future. Thus, interactions tend to be tedious and repetitive. By adding a personalized, long-term user model, the quality of these interactions can improve drastically. At the same time, collecting user preferences in recommendation systems often requires form filling or other explicit statements of preferences on the user's part, which can be difficult and time consuming. Collecting preferences in the course of the dialogue lets the user begin the task of item search immediately.

The interaction between conversation and personalized recommendation has also affected our choices for the acquisition, utilization, and representation of user models. The ADAPTIVE PLACE ADVISOR learns information about users unobtrusively, in the course of a normal conversation whose purpose is to find a satisfactory item. The system stores this information for use in future conversations with the same individual. Both acquisition and utilization occur not only when items are presented to and chosen by the user, but also during the search for those items. Finally, the system's representation of models goes beyond item preferences to include preferences about both item characteristics and particular values of those characteristics. We believe that these ideas extend to other types of preferences and other types of conversations.

In this paper, we describe our work with the ADAPTIVE PLACE ADVISOR. We begin by introducing personalized and conversational recommendation systems, presenting our design decisions along the way. In Section 3 we describe the system in detail, while in Section 4 we present our experimental evaluation. In Sections 5 and 6 we discuss related and future work, respectively, and in Section 7 we conclude and summarize the paper.

## 2. Personalized Conversational Recommendation Systems

Our research goals are two-fold. First, we want to improve both interaction quality in recommendation systems and the utility of results returned by making them user adaptive and conversational. Second, we want to improve dialogue system performance by means of personalization. As such, our goals for user modeling differ from those commonly assumed in recommendation systems, such as improving accuracy or related measures like precision and recall. Our goals also differ from that of previous work in user modeling in dialogue systems (Haller & McRoy, 1998; Kobsa & Wahlster, 1989; Carberry, 1990; Kass, 1991), which emphasizes the ability to track the user's goals as a dialogue progresses, but which does not typically maintain models across multiple conversations.

Our hypothesis is that improvements in efficiency and effectiveness can be achieved by using an unobtrusively obtained user model to help direct the system's conversational search for items to recommend. Our approach assumes that there is a large database of items from which to choose, and that a reasonably large number of attributes is needed to describe these items. Simpler techniques might suffice for situations where the database is small or items are easy to describe.

### 2.1 Personalization

Personalized user adaptive systems obtain preferences from their interactions with users, keep summaries of these preferences in a user model, and utilize this model to generate customized information or behavior. The goal of this customization is to increase the quality and appropriateness of both the interaction and the result(s) generated for each user.

The user models stored by personalized systems can represent stereotypical users (Chin, 1989; Rich, 1979) or individuals, they can be hand-crafted or learned (e.g., from questionnaires, ratings, or usage traces), and they can contain information about behavior such as previously selected items, preferences regarding item characteristics (such as location or price), or properties of the users themselves (such as age or occupation) (Kobsa & Wahlster, 1989; Rich, 1979). Also, some systems store user models only for the duration of one interaction with a user (Carberry, Chu-Carroll, & Elzer, 1999; Smith & Hipp, 1994), whereas others store them over the long term (Rogers, Fiechter, & Langley, 1999; Billsus & Pazzani, 1998).

Our approach is to learn probabilistic, long-term, individual user models that contain information about preferences for items and item characteristics. We chose learned models due to the difficulty of devising stereotypes or reasonable initial models for each new domain encountered. We chose probabilistic models because of their flexibility: a single user can exhibit variable behavior and their preferences are relative rather than absolute. Long-term models are important to allow influence across multiple conversations. Also, as already noticed, different users have different preferences, so we chose individual models. Finally, preferences about items and item characteristics are needed to influence conversations and retrieval.

Once the decision is made to learn models, another design decision relates to the method by which a system collects preferences for subsequent input to the learning algorithm(s). Here we can distinguish between two approaches. The *direct feedback* approach places the burden on the user by soliciting preference information directly. For example, a system might ask the user to complete a form that asks her to classify or weight her interests using a variety of categories or item characteristics. A recent study (McNee, Lam, Konstan, & Riedl, 2003) showed that forcing the user to provide ratings for items (movies, in this case) that they choose, rather than those that the system chooses, can actually lead to better accuracy rates and better user loyalty. However, users can be irritated by the need to complete long questionnaires before they can even begin to enjoy a given service, and the study was not in the context of a dialogue system but involved a simpler interaction. Another, slightly less obtrusive, form of direct feedback encourages the user to provide feedback as she continues to use a particular service.

The second approach to acquiring user models, and the one taken in the Adaptive Place Advisor, is to infer user preferences *unobtrusively*, by examining normal online behavior (Fiechter & Rogers, 2000; Rafter, Bradley, & Smyth, 2000). We feel that unobtrusive collection of preferences is advantageous, as it requires less effort from the user. Also, users often cannot articulate their preferences clearly until they learn more about the domain. A possible disadvantage to unobtrusive approaches is that users may not trust or understand the system's actions when they change from one interaction to the next. This could be addressed by also letting the user view and modify the user model (Kay & Thomas, 2000).

Systems typically take one of two approaches to preference determination. *Content-based* methods recommend items similar to ones that the user has liked in the past (Segal & Kephart, 1999; Pazzani, Muramatsu, & Billsus, 1996; Lang, 1995). In contrast, *collaborative* methods select and recommend items that users similar to the current user have liked in previous interactions (Cotter & Smyth, 2000; Billsus & Pazzani, 1998; Konstan, Miller, Maltz, Herlocker, Gordon, & Riedl, 1997; Shardanand & Maes, 1995). Because collaborative filtering bases recommendations on previous selections of other users, it is not suitable for new or one-off items or for users with uncommon preferences. The content-based approach, on the other hand, uses the item description itself for recommendation, and is therefore not prone to these problems. However, content-based techniques tend to prefer the attribute values that users have preferred in the past, though they do allow new combinations of values. We feel that the benefits of a content-based approach outweigh the disadvantages; we discuss methods for overcoming these disadvantages and for combining the two techniques in Section 6.3.

Ultimately, personalization is about how one can utilize a learned user profile to search for, identify, and present relevant information to the right user in the right way at the right time. User models have been utilized in recommendation systems for content processing and selection (information filtering), navigation support in web browsers (Pazzani et al., 1996), and choice of modality and style of presentation and interaction (Brusilovsky & Maybury, 2002). The Adaptive Place Advisor adapts its information filtering and interaction behavior, since these are most relevant for our application and since the majority of the interaction is through natural language.

## 2.2 Conversational Recommendation

The goal of recommendation systems is to help users find items of interest, and thus they must eliminate some choices from consideration while keeping others. To achieve this, they must either know or determine the interests of the current user, then present items that satisfy those interests.

One of the most widely used approaches for information acquisition in recommendation systems is the use of a simple form (usually a single query box) that the user fills in. Coupled with this, a common presentation format is the *ranked list* approach. In this scheme, the system presents a usually long list of results, ordered by their predicted relevance to the user. The combined technique of information acquisition through a form and presentation in a list originated in information retrieval, where it was used to help find documents or reports during library research. The basic method, which we refer to as the ranked list approach, is more general, however, and it underlies the simpler search engines for the World Wide Web, which are used regularly for many different selection tasks.

In contrast, the ADAPTIVE PLACE ADVISOR supports *conversations* as its mode of interaction. In this approach, one conversational participant, which we call the *inquirer*, asks a second participant, which we call the *advisor*, for assistance in selecting an item. For example, if the inquirer is deciding on a restaurant at which to have dinner, the conversation may proceed as follows:

1. Inquirer: Where do you think I should eat tonight?
2. Advisor: What type of food would you like?
3. Inquirer: What types are there?
4. Advisor: You can say things like Chinese, Indian, and Mediterranean.
5. Inquirer: Oh, maybe a cheap Indian place.
6. Advisor: What kind of parking would you like?
7. Inquirer: I don't care, as long as it's in Palo Alto.
8. Advisor: I'm sorry, I don't know of any restaurants like that, would you like to search for any price?
9. Inquirer: No, I think I'd like Chinese instead.
10. Advisor: How does Mandarin Gourmet on 420 Ramona sound?[1]
11. Inquirer: No, what else do you have?
12. Advisor: How about Jing-Jing Szechwan Hunan Gourmet on 443 Emerson Street?
13. Inquirer: Sure, that sounds fine.

We will refer to systems that mimic this approach to recommendation as *conversational recommendation systems.*

The interaction supported by conversational recommendation systems is quite different from that found in the ranked list approach. The most important distinction is that the interaction takes the form of a sequence of questions, most designed to eliminate some items from consideration. Answering these questions plays a similar role to giving keywords in a form, but the aim is to remove alternatives rather than to rank results in a list. This conversational process can also help the inquirer better understand his own desires, since thinking about possible questions and answers may clarify goals in ways that keyword entry does not encourage. In addition, a conversation can support user variability by letting the user control item and attribute choices in each conversation. Finally, the inquirer never hears about a complete item until at most a few choices remain. This avoids presenting a long list of items and lets the user narrow down the choices in an iterative, manageable fashion.

---

1. Our system does not support this exact sample dialogue. For example, at this point in a conversation it displays the information about the restaurant on the screen and asks "How does this one sound?" or a similar question.

Such dialogues seem better for recommendations that must be delivered by speech rather than visually, for example those engaged in while the inquirer is driving. They also seem ideal, independent of modality, for tasks like destination selection or help-desk support (Göker & Roth-Berghofer, 1999; Aha & Breslow, 1997), in which the user needs to converge on at most a few items. On the other hand, keyword entry and ranked list methods seem more appropriate in situations where the user prefers to provide all requirements at once, in situations where information can be presented visually, and in situations where the user may want to examine many options.

By eliminating options, conversational recommendation systems ultimately direct their users to a suitable solution. However, such a conversation can become tiring and the quality of the first result returned may not be acceptable for each user. Just as interactions with a friend who knows your concerns can be more directed and produce better results than those with a stranger, dialogues with a conversational advisor should become more efficient and effective over time. Our goals for user modeling include improvement of the subjective quality and effectiveness of both the results (found items) and the conversation that leads to these results. For example, after several conversations with the inquirer above, a new interaction may proceed as follows, where the question about parking has been eliminated and the item presentation order has changed:

1. Inquirer: Where do you think I should eat tonight?
2. Advisor: What type of food would you like?
3. Inquirer: Oh, maybe a Chinese place.[2]
4. Advisor: What city do you prefer?
5. Inquirer: Do you have something in Palo Alto?
6. Advisor: How does Jing-Jing Szechuan Gourmet on 443 Emerson sound?
7. Inquirer: Sure, that sounds fine.

We turn next to our design choices concerning the management of such conversations.

## 2.3 Conversation via Dialogue Management

Dialogue systems carry out conversations with users in natural language, whether spoken or typed. The main tasks performed by dialogue systems are language interpretation, language generation, and dialogue management. Natural language interpretation and generation are topics onto themselves and we will not discuss them here; for two introductory texts, see Allen (1995) and Jurafsky and Martin (2000). To enable a focus on user modeling, our system allows moderately complex user utterances but has a pre-coded set of system utterances, as discussed further in Section 3.3.

---

2. This response shows that the Inquirer will have learned how to use the system more efficiently as well.

The simplest dialogue managers are based on finite-state automata in which the states correspond to questions and arcs correspond to actions that depend on a user-provided response (Stent, Dowding, Gawron, Bratt, & Moore, 1999; Winograd & Flores, 1986). These systems support what are called fixed- or system-initiative conversations, in which only one of the participants controls the actions, whether it be the system helping the user or the user asking questions of the system. Next in complexity are frame- or template-based systems in which questions can be asked and answered in any order (Bobrow et al., 1977). Next, true mixed-initiative systems allow either dialogue participant to contribute to the interaction as their knowledge permits (Allen, 1999; Haller & McRoy, 1998; Pieraccini, Levin, & Eckert, 1997). Thus, the conversational focus can change at any time due to the user's (or system's) initiative of that change. Finally, some different approaches that support sophisticated dialogues include plan-based systems (Allen et al., 1995; Cohen & Perrault, 1979) and systems using models of rational interaction (Sadek, Bretier, & Panaget, 1997).

To allow reasonably complex conversations while keeping the system design straight-forward, we chose a frame-based approach to dialogue management. Thus, the Adaptive Place Advisor allows more conversational flexibility than a fully system-initiative paradigm would allow. Users can fill in attributes other than or in addition to those suggested by the system. However, they cannot force the system to transition to new subtasks, nor can the system negotiate with users to determine which participant should take the initiative.

## 2.4 Interactive Constraint-Satisfaction Search

Constraint-satisfaction problems provide a general framework for defining problems of interest in many areas of artificial intelligence, such as scheduling and satisfiability (Kumar, 1992). In their most general form, constraint-satisfaction problems involve a set of variables whose domains are finite and discrete, along with a set of constraints that are defined over some subset of the variables and that limit the value combinations those variables can take. The goal is to find an assignment of values to variables that satisfies the given constraints.

Cucchiara, Lamma, Mello, and Milano (1997) define the class of *interactive* constraint-satisfaction problems that involve three extensions to the standard formulation. First, they include a constraint acquisition stage during which a user can incrementally add new constraints to the problem being solved. Second, a variable's domain can include both a defined and undefined portion, and the user can add new values to the defined portion during constraint acquisition. Third, they allow incremental update of a partial solution based on the domain and constraint updates.

This framework can encompass the item search portion of the conversations managed by the Adaptive Place Advisor; it does not include the item presentation portion. In our setting, constraints are simply attribute-value specifications, such as *cuisine = Chinese*. The Place Advisor's search is not as fully general as this framework, in that it does not incorporate the notion of undefined portions of domains. However, it does acquire constraints via the user's specifications during a conversation and incrementally updates solutions in response.

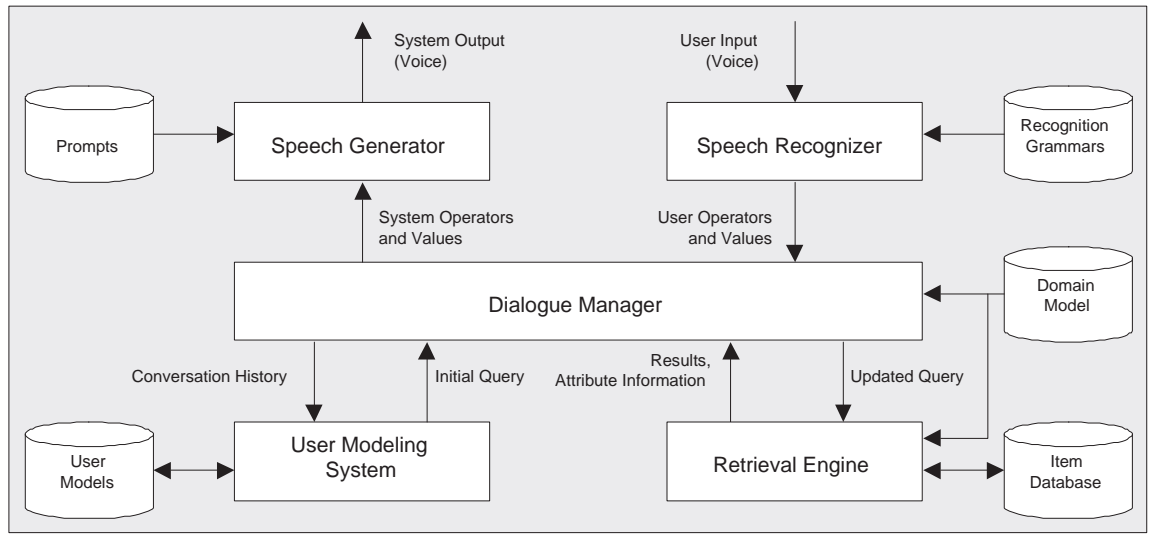

Figure 1: Components of the Adaptive Place Advisor and their interactions.

## 3. The Adaptive Place Advisor

In this section, we first present an overview of the Adaptive Place Advisor's functioning, then follow with details about its components.[3] The system carries out a number of tasks in support of personalized interaction with the user; in particular, it:

- utilizes the user model to initialize a probabilistic item description as an *expanded query*,
- generates context-appropriate utterances,
- understands the user's responses,
- refines the expanded query with the explicit requirements (constraints) obtained from the user during the conversation,
- retrieves items matching the explicitly specified part of the query from a database,
- calculates the similarity of the retrieved items to the query,
- selects the next attribute to be constrained or relaxed during a conversation when the number of highly similar items is not acceptable,
- presents suitable items when the number of items is acceptable, and
- acquires and updates the user model based on these interactions.

The responsibilities for these tasks are distributed among various modules of the system, as shown in Figure 1. The Dialogue Manager generates, interprets, and processes conversations; it also updates the expanded query after each user interaction. The Retrieval Engine is a case-based reasoning system (Aamodt & Plaza, 1994) that uses the expanded query to retrieve items from the database and to measure their similarity to the user's preferences. The User Modeling System generates the initial (probabilistic) query and updates the long-term user model based on the conversation history. The Speech Recognizer and the Speech Generator handle the user's input and control the system's output, respectively.

---

3. As further discussed in Section 5.2, our approach to destination advice draws on an earlier analysis of the task by Elio and Haddadi (1998, 1999).

To find items to recommend to the user, the PLACE ADVISOR carries out an augmented interactive constraint-satisfaction search. The goal of the entire conversation is to present an item that will be acceptable to the user. During the constraint-satisfaction portion, the system carries out a conversation to find a small set of such items. During the search phase, two situations determine the system's search operators and thus its questions. First, an under-constrained specification means that many items match the constraints, and the system must obtain more information from the user. Second, if there are no matching items, the system must relax a constraint, thus allowing items to contain any domain value for the relaxed attribute.[4] The system ends the search phase when only a small number of items match the constraints and are highly similar (based on a similarity threshold) to the user's preferences. Item presentation (in similarity order) begins at this point, with a similarity computation used to rank the items that satisfy the constraints.

The search and item presentation process is also influenced by the User Modeling System and thus is *personalized*. The main mechanism for personalization is through the expanded query, a probabilistic representation of the user's preferences, both long-term (over many conversations) and short-term (within a conversation). We will often just refer to this as the "query," but it always refers to constraints that are both explicitly and implicitly specified by the user. Thus, the query is "expanded" beyond the explicit (short-term) constraints using the (long-term) constraints implicit in the user model. In a sense, the initial query represents what constraints the system thinks the user will "probably want." The system incrementally refines this query in the course of the conversation with the user, setting explicit, firm constraints as the user verifies or disconfirms its assumptions. Over the long term, the User Modeling System updates the user model based on the user's responses to the attributes and items offered during a conversation.

The Retrieval Engine searches the database for items that match the explicit constraints in the query. It then computes the similarity of the retrieved items to the user's preferences as reflected in the expanded part of the query. Depending on the number of highly similar results, the Retrieval Engine also determines which attribute should be constrained or relaxed.

In sum, the system directs the conversation in a manner similar to a frame-based system, retrieves and ranks items using a case-based reasoning paradigm, and adapts the weights in its similarity calculation based on past conversations with a user, thereby personalizing future retrievals and conversations. In this section, we present the details of the ADAPTIVE PLACE ADVISOR's architecture. After describing the user model, we elaborate on the Retrieval Engine then the Dialogue Manager. Finally, we discuss how the system updates the user model as the user interacts with it.

## 3.1 The User Model

Our focus on personalized conversation suggests a fine-grained model of user preferences, emphasizing the questions a user prefers to answer and the responses he tends to give, in addition to preferences about entire items. We now describe that model in more detail. In later sections, we will describe how it influences item ranking and question ordering, which in turn determine how quickly the system can stop asking questions and start presenting items. In general, a user may tend to:

- answer questions about some *attributes* more often than others,
- provide some attribute *values* more often than others,
- choose some *items* more often than others,
- provide certain *combinations* of values more often than their independent distribution would predict, and
- accept either large or small amounts of value and item *diversity*.

---

4. Because other constraints can later be modified, the system lets the user later specify any value, even the one that caused the over-constrained situation.

Table 1: Example of a user model.

| User Name | | Homer | | | | | |
|---|---|---|---|---|---|---|---|
| Attributes | $w_i$ | Values and probabilities | | | | | |
| Cuisine | 0.4 | Italian | French | Turkish | Chinese | German | English |
| | | 0.35 | 0.2 | 0.25 | 0.1 | 0.1 | 0.0 |
| Price Range | 0.2 | one | two | three | four | five | |
| | | 0.2 | 0.3 | 0.3 | 0.1 | 0.1 | |
| ... | ... | ... | | | | | |
| Parking | 0.1 | Valet | | Street | | Lot | |
| | | 0.5 | | 0.4 | | 0.1 | |
| Item Nbr. | | 0815 | 5372 | 7638 | ... | | 6399 |
| Accept/Present | | 23 / 25 | 10 / 19 | 33 / 36 | ... | | 12 / 23 |

All of these tendencies are influenced by the user's preferences, which in turn are captured by our user model. Attribute preferences represent the relative importance a user places on attributes (e.g., cuisine vs. price) while selecting an item. Preferred values show the user's bias towards certain item characteristics (e.g., Italian restaurants vs. French restaurants). Item preferences are reflected in a user's bias for or against a certain item, independent of its characteristics. Combination preferences represent constraints on the combined occurrence of item characteristics (e.g., accepts restaurants in San Francisco only if they have valet parking). Diversity preferences model the time that needs to pass between an item or characteristic being suggested again or the user's tolerance for unseen values or items. Item preferences are related to single items, whereas attribute, value, and combination preferences are applicable to the search for those items in general. Diversity preferences relate to both the items and the search.

Currently, the ADAPTIVE PLACE ADVISOR models preferences that the user may have about attributes, values, and items, but not combination or diversity preferences. The former are easily captured by either probability distributions or counts, as illustrated in Table 1. The PLACE ADVISOR maintains a probability distribution to represent attribute preferences and independent probability distributions to represent preferences for each attribute's set of values. For attribute preferences, the system uses domain knowledge to initialize the weights; for example, price is usually considered as more important than park-ing. In the absence of such information, as is the case with value preferences, the system begins with a uniform distribution.

The system represents item preferences as a ratio of the number of times an item was accepted to the number of times it was presented; this is initialized by assuming that all items have been presented and then accepted a large percentage (nine out of ten, or 90%) of the time. While this may cause updates (see below) to have a small effect and undesirable items to be suggested more than once, it has the effect of not quickly discounting alternatives early in the learning process. This in turn encourages the user to explore alternatives, allowing the system to learn more about additional items. In sum, item preferences represent the probability of the user accepting a particular item after it is presented, rather than representing a probability distribution over all items.

## 3.2 The Retrieval Engine

Once in place, the user model affects the behavior of the system through the Retrieval Engine, which interacts with a database to retrieve the items, if any, that satisfy the currently agreed upon constraints. This module also interacts with the query to determine how similar those items are to the user's preferences and to determine the best ordering of the attributes for constraining or relaxing, as appropriate. Both types of interactions with the user model support the goal of quickly narrowing down the search for a satisfactory item.

Similar to the way a human advisor bases assumptions regarding the inquirer on their previous interactions, our system uses its cumulative experience, reflected in the user model, as a basis for its computation. The Retrieval Engine represents the user's preferences and requirements in the expanded query, a partial, probabilistic item specification determined by both the conversation and the user model. The query is initialized from the user model, and thus contains preference-based probabilities for the attributes and values the user has not yet explicitly specified along with the user's item preferences. In the course of the conversation, the system updates the query to reflect the values the user specifies. For each attribute, it sets the "probability" for each value agreed upon in the conversation to 1.0, and all other probabilities to zero. For example, if the user says "Chinese or Italian," the system sets the value probabilities for both Chinese and Italian to 1.0, and all other cuisine probabilities to zero. This is equivalent to a disjunction of probabilistic queries, one for each value combination specified by the user.[5]

The first main aspect of the Retrieval Engine that is personalized is its item ranking technique. Unlike a typical case-based similarity computation, which retrieves items beyond those that match a query exactly, the computation used by the system restricts the retrieved items to those most desirable to the user. The system filters the items to be included in the current case base according to the characteristics explicitly specified by the user and sorts the remaining items according to their similarity to items the user liked in the past.

Thus, the Engine first queries the database to retrieve all items that match the current constraints exactly.[6] Then, the PLACE ADVISOR uses the probability distributions in the query to compute how likely the user is to choose an item. The system calculates the

---

5. In the user study described in Section 4, no users specified a disjunctive query.

6. For attributes where the user has selected more than one value, we assume that any supplied value would be acceptable.

similarity between the current query, $Q$, and an item, $I$ using

$$Sim(Q, I) = R_I \times \sum_{j=1}^{n} w_j \times P(V_j) \, ,$$

where $R_I$ is the user's item preference ratio for item $I$, $n$ is the number of attributes, $w_j$ is the weight of attribute $j$ in $Q$, $V_j$ is the value of attribute $j$ in $I$, and $P(V_j)$ is the value preference (probability in $Q$) for this value. Similarity in this formula is based on the user model and the search state. Thus, for each unconstrained attribute, it estimates the probability that the user will accept the item's value for that attribute. Once the system calculates the similarity of each item, it compares an item's similarity to a constant similarity threshold, and only retains those items that exceed the threshold.

The second main personalized aspect of the Retrieval Engine is its ranking of attributes in under- and over-constrained situations. This helps assure that the user is more likely to respond informatively to the system's questions in under-constrained situations, and to allow the suggested attribute to be relaxed in over-constrained situations. For both situations, one option is to order attributes randomly, which is a technique used in some simple dialogue systems. Another option is to use a conditional entropy measure to select attributes (see Section 6.1). A third is to rank attributes in order by their desirability to the user, as reflected in the user model, and we take this option.

In addition to using the long term user model to rank attributes, the system also uses the attribute weights that are reflected in the query. We will see see later that the query's attribute weights, while initialized from the user model, are also influenced by the conversation. In an over-constrained situation, the attribute ranking order is from highest to lowest, and in an under-constrained situation, it is the reverse.[7] Using the attribute weights rather than the conditional entropy avoids pitfalls that arise from the continuously changing value distribution in the data (new restaurants, restaurants going out of business, etc.). If this did affect attribute rankings, the users may be confused by the resulting variability. But even worse, it would not reflect the user's preferences. Further, not every question that has a high information gain is of high relevance for a user selecting a destination (e.g., "parking options" may have a very high score but should only be asked after the user has decided on cuisine and location.)

In summary, the user model influences item retrieval, item ranking, and attribute ranking, which in turn influence the system's utterances during the conversation.

### 3.3 Conversing with the User

As it converses with the user, the Dialogue Manager uses the results of the Retrieval Engine's functions. The system uses a frame containing a simple list of constraints to support the interactive constraint-satisfaction search (see Jurafsky et al. 1994 or Dowding et al. 1993 for a similar formulation). As is usual in this type of system, the user can respond to a system request to fill a constraint by ignoring that attribute and specifying the value for different one(s) instead (Goddeau et al., 1996; Ward & Issar, 1996). The speech acts supported are listed in Table 2.

---

7. CBR systems do not necessarily use the same weighting factors for each of similarity computation and question ordering. However, for our application area, it is correct to make the assumption that an attribute's importance is the same as its impact on the similarity computation.

Table 2: Speech acts supported in the Adaptive Place Advisor.

| **System Speech Acts** | |
|---|---|
| Attempt-Constrain | Asks a question to obtain a value for an attribute. |
| Suggest-Relax | Asks a question to remove all values for an attribute. |
| Recommend-Item | Recommends an item that satisfies the constraints. |
| Quit-Start-Mod | States that no matching items remain and asks whether to modify the search, start over, or quit. |
| Provide-Values | Lists a small set of values for an attribute. |
| Clarify | Asks a clarifying question. |
| **User Speech Acts** | |
| Provide-Constrain | Provides a value for an attribute. |
| Accept | Accepts a relaxation suggestion or item generated by the system. |
| Reject | Rejects the system's proposed attribute, relaxation attempt, or item. |
| Provide-Relax | Provides an attribute value for removal. |
| Start-Over | Indicates a desire to reinitialize the constraints and begin again. |
| Quit | Indicates a desire to stop the conversation. |
| Query-Values | Asks for information about possible values of an attribute. |

Table 3: Dialogue State.

| *Variable* | *Description* |
|---|---|
| `Constrained` | Attributes whose values have been specified. |
| `Rejected` | Attributes whose value the user has declined to provide. |
| `Fixed` | Constrained attributes that the user has indicated should not be relaxed. |
| `Constrain` | The next attribute to constrain, if any. |
| `Relax` | The next attribute to relax, if any. |
| `Query` | Probability model of desired item constraints. |
| `Number-of-Items` | Number of database items matching the query and exceeding the similarity threshold. |
| `Ranked-Items` | The matching items ranked in similarity order. |
| `Rejected-Items` | Items that the user has rejected. |
| `User-Move` | The user's most recently uttered speech act. |
| `System-Act` | The system's most recently uttered speech act. |

There are two main phases of the dialogue, the interactive constraint-satisfaction portion and the item presentation portion. The constraint-satisfaction portion is further divided into over- and under-constrained situations. The dialogue state (Table 3) determines the system's utterance and the range of responses expected at each point. The system updates the dialogue state's variables as appropriate throughout the conversation.

Table 4: Sample Conversation.

1. Inquirer: Where do you think I should eat tonight?
2. Advisor: What type of food would you like?
3. Inquirer: What types are there?
4. Advisor: You can say things like Chinese, Indian, and Mediterranean.
5. Inquirer: Oh, maybe a cheap Indian place.
6. Advisor: What kind of parking would you like?
7. Inquirer: I don't care, as long as it's in Palo Alto.
8. Advisor: I'm sorry, I don't know of any restaurants like that, would you like to search for any price?
9. Inquirer: No, I think I'd like Chinese instead.
10. Advisor: How does Mandarin Gourmet on 420 Ramona sound?
11. Inquirer: No, what else do you have?
12. Advisor: How about Jing-Jing Szechwan Hunan Gourmet on 443 Emerson Street?
13. Inquirer: Sure, that sounds fine.

In more detail, the system's speech act (or *move*) during interactive constraint-satisfaction is determined by the `Number-of-Items` dialogue state variable. Further, its speech act determines which speech recognition grammar to employ to interpret the user's next utterance. The most common situation is when many items (more than some small threshold, here three) match the current constraints.[8] In this situation, the system makes an Attempt-Constrain move, in which it asks the user to fill in the value for an attribute. This move, if responded to appropriately by the user, would reduce the number of items considered to be satisfactory to the user. The attribute to `Constrain` is the one ranked highest by the Retrieval Engine that has not already been `Constrained` or `Rejected`. In our first sample conversation, repeated in Table 4, utterances 2 and 6 illustrate Attempt-Constrain's by the system.

One user response to an Attempt-Constrain is a Provide-Constrain, in which he provides a value for the specified attribute or for additional attributes, as in utterances 5 and 7. A second possible response is a Reject, in which the user indicates disinterest or dislike in an attribute, as in the first part of utterance 7. As illustrated by some of these examples, the user can combine more than one move in a single utterance.

A second situation, an *over-constrained* query, occurs when there are no items that satisfy the agreed upon constraints and are similar enough to the user's preferences, and thus the Retrieval Engine returns an empty set (`Number-of-Items = 0`). In this case, the system performs a Suggest-Relax move that informs the user of the situation and asks if he would like to relax a given constraint. The attribute to `Relax` is chosen from the Retrieval Engine's highest ranked attribute[9] that has not already been `Fixed`. This is illustrated in utterance 8 of the conversation in Table 4. As in utterance 9 of that conversation, the user can respond by rejecting (Reject) the system's suggestion or he can accept it (Accept). In the former case, the attribute is `Fixed` so that the system will not try to relax it again.

---

8. When we discuss the number of items matching the constraints, we refer to those items that remain after similarity filtering as discussed in Section 3.2.

9. Recall from Section 3.2 that this is actually the lowest ranking attribute in the user model.

In combination with either of these speech acts, the user can specify other attributes to relax in addition to, or instead of, the system-suggested attribute (Provide-Relax).

When only a few items satisfy the constraints, the system ends the interactive search and begins to suggest items to the user (Recommend-Item) in order of similarity, as in utterances 10 and 12 above. The user can either accept or reject an item. If the user accepts an item (Accept), the system ends the conversation, having reached the goal state. If the user rejects an item (Reject), the system presents an alternative, if any remain. Note that there are three "meanings" for the Reject speech acts of the user, but only two "meanings" for the Accept speech acts, since a user has to accept an Attempt-Constrain by providing an explicit value for the attribute being constrained.

There are three special situations not covered by the above. The first is when the query is over-constrained, but the user has `Fixed` all attributes that could be relaxed. The second is when the user has rejected all items that match the constraints. In these two situations, the system informs the user of the situation, asks him whether he would like to quit, start over, or modify the search (Quit-Start-Mod), and reacts accordingly. The third special situation is when `Number-of-Items` exceeds the presentation threshold, but all attributes have been `Constrained` or `Rejected`. In that case, the Place Advisor begins to present items to the user.

To support the spoken natural language input and output, we use a speech recognition package from Nuance Communications, Inc. This package lets us write a different recognition grammar for each of the situations described above and to use human-recorded prompts (rather than text-to-speech). The string of words recognized by the system is parsed using recognition grammars that we wrote, which were used for all users without adaptation. Future work could include personalized recognition grammars as well as personalized information preferences. The grammars use semantic tags to fill in each slot: besides slots for each attribute, we define slots for rejection or acceptance of the system's suggestions. In more complex domains, more sophisticated parsing methods may be required, but this simple scheme gives the user a reasonably diverse set of utterance options. The Nuance modules also generate a response to user requests for help (Query-Values) with a Provide-Values speech act, and enter clarification dialogues when the confidence in a recognized utterance is below a given threshold (Clarify). These are currently simple interactions where the system provides examples of answers to the most recently uttered prompt, or asks the user to repeat themselves.

Finally, for the item presentation portion of the dialogue only, the system displays the restaurant information (name, address, and phone number) on the screen, and outputs a spoken prompt such as "How about this one?" We chose this presentation modality due to our reluctance to use text-to-speech generation and the large number of prompts we would have had to record to produce spoken language for each restaurant. However, note that the user still responds with a spoken reply, and we do not feel that this presentation mode substantially influenced the user-modeling behavior of the Place Advisor.

Each system-user interaction affects subsequent rounds of database retrieval and similarity calculation via updates to the expanded query. Table 5 shows the effects of relevant speech acts on the query, which is in turn used in the similarity calculation as described in Section 3.2. In the table, we have shortened the names of some of the system moves for the sake of brevity.

Table 5: The effects of speech acts on the query. Key: Constrain = Attempt-Constrain, Relax = Suggest-Relax, Recommend = Recommend-Item.

| System-Move | User-Move | Effect on Query |
|---|---|---|
| Constrain | Provide-Constrain | Set "probabilities" of all provided values to one. Set probability of other values for constrained attributes to zero. If the attribute has been rejected previously, reset its attribute probability from user model. |
| Constrain | Reject | Drop attribute by setting its probability to zero. |
| Relax | Reject | No effect; Dialogue Manager selects next attribute. |
| Recommend | Reject | Update item preference counts (see Section 3.4). |
| Relax | Accept | Reset value probabilities for the attribute from user model. |
| Recommend | Accept | Update item preference counts (see Section 3.4). |
| Any | Provide-Relax | Reset value probabilities for the attribute from user model. |
| Any | Start-Over | Initialize from user model. |

## 3.4 Updating the User Model

Our main contribution is the addition of personalization to the above conversational recommendation model. The user model (Section 3.1) represents this personalization, but the Adaptive Place Advisor must update it appropriately. While some adaptive recommendation systems (Smyth & Cotter, 1999; Linden, Hanks, & Lesh, 1997; Pazzani et al., 1996; Lang, 1995) require the user to provide direct feedback to generate the user model, our basic approach is to unobtrusively derive the user preferences. Thus, the system does not introduce unnecessary interactions, but learns from the interactions needed to support item recommendation. We describe here how the system gathers item, attribute, and value preferences. As described more fully below, the system modifies feature and value weights (Fiechter & Rogers, 2000; Zhang & Yang, 1998; Bonzano, Cunningham, & Smyth, 1997; Wettschereck & Aha, 1995) for the latter two, and increases the counts in the ratio of accepted to presented items.

When determining the points in the dialogue at which to update the user model, we considered several factors. We wanted to enable the system to acquire information quickly, but to discourage it from making erroneous assumptions. We thought that users might explore the search space the most while constraining attributes, so we decided not to have the system update value preferences after each user-specified constraint. However, if we instead chose to only allow model updates after an item suggestion, the learning process might be too slow. The choices described below are, we feel, a good tradeoff between the extremes.

The three circumstances that we chose for user model update were (1) after the user's Accept speech acts in a Suggest-Relax situation, (2) after the user's Accept speech acts in a Recommend-Item situation, and (3) after the user's Reject speech act after a Recommend-Item speech act by the system. First, we assume that when a user accepts an item, he is indicating: (1) a preference for the item itself, (2) preferences for the attributes he constrained to find this item, and (3) preferences for the values he provided for those attributes. Thus, when a user accepts an item presented by the system, the probabilities for the appropriate item, attributes, and values are increased. For the item preference, the system simply adds one to the presentation and acceptance counts. For attribute and value preferences, the system increases the probability of the appropriate weight by a small amount proportional to its current weight, then renormalizes all weights. Thus attribute and value preferences are biased measures that avoid zero counts for values that the user never chooses, as is typical for this type of probabilistic representation.

Second, when a user rejects an item presented by the system, we only assume that he has a dislike for the particular item. We do not assume anything about the characteristics of that item, since the user has specified some of those characteristics. Therefore, for rejected items the system simply adds one to the presentation count.

The third situation in which the system updates the user model is when, after the query has become over-constrained, it presents an attribute for relaxation and the user accepts that relaxation. In this situation, we assume that, had there been a matching item, the user would have been satisfied with it, since the characteristics specified in the conversation so far were satisfactory. Therefore, before the relaxation occurs, the system increases the attribute preferences for the constrained attributes and increases the value preferences for user-specified values, in a manner similar to an ACCEPT situation after a RECOMMEND-ITEM. This enables the ADAPTIVE PLACE ADVISOR to more quickly make inferences about the user's preferences.

## 4. System Evaluation

As stated earlier, we believe that user modeling increases the effectiveness and efficiency of conversations with the system over time. To test this hypothesis, we carried out an experiment with a version of the ADAPTIVE PLACE ADVISOR that recommends restaurants in the San Francisco Bay Area. The system describes items using seven attributes: cuisine, rating, price, location, reservations, parking options, and payment options. Most attributes have few values, but cuisine and location have dozens. There are approximately 1900 items in the database.

We asked several users, all from the Bay Area, to interact with the system to help them decide where to go out to eat. The users were given no external guidance or instructions on which types of restaurants to select, other than to look for and choose those that they might actually patronize. An experimenter was present during all these interactions, which were filmed, but his help was not needed except on rare occasions when a subject repeatedly tried words that were not included in the speech recognition grammar.

### 4.1 Experimental Variables

To test our hypothesis about the benefits of personalization in the ADAPTIVE PLACE ADVISOR, we controlled two independent variables: the presence of user modeling and the number of times a user interacted with the system. First, because we anticipated that users might improve their interactions with the PLACE ADVISOR over time, we divided subjects into an experimental or modeling group and a control group. The 13 subjects in the modeling group interacted with a version of the system that updated its user model as described in Section 3.4. The 11 subjects in the control group interacted with a version that did not update the model, but that selected attributes and items from the default distribution described in Section 3.1. Naturally, the users were unaware of their assigned group. Second, since we predicted the system's interactions would improve over time, as it gained experience with each user, we observed its behavior at successive points along this "learning curve." In particular, each subject interacted with the system for around 15 successive sessions. We tried to separate each subject's sessions by several hours, but this was not always possible. However, in general the subjects did use the system to actually help them decide where to eat either that same day or in the near future; we did not provide constraints other than telling them that the system only knew about restaurants in the Bay Area.

To determine each version's efficiency at recommending items, we measured several conversational variables. One was the average number of *interactions* needed to find a restaurant accepted by the user. We defined an interaction as a cycle that started with the system providing a prompt and ended with the system's recognition of the user's utterance in response, even if that response did not answer the question posed by the prompt. We also measured the *time* taken for each conversation. This began when a "start transaction" button was pushed and ended when the system printed "Done" (after the user accepted an item or quit).

We also collected two statistics that should not have depended on whether user modeling was in effect. First was the number of *system rejections*, that is, the number of times that the system either did not obtain a recognition result or that its confidence was too low. In either case the system asked the user to repeat himself. Since this is a measure of recognition quality and not the effects of personalization, we omitted it from the count of interactions. A second, more serious problem was a speech *misrecognition* error in which the system assigned an utterance a different meaning than the user intended.

Effectiveness, and thus the subjective quality of the results, was somewhat more difficult to quantify. We wanted to know each user's degree of satisfaction with the system's behavior. One such indication was the *rejection rate*: the proportion of attributes about which the system asked but the subject did not care (REJECT's in ATTEMPT-CONSTRAIN situations). A second measure was the *hit rate*: the percentage of conversations in which the first item presented was acceptable to the user. Finally, we also administered a questionnaire to users after the study to get more subjective evaluations.

## 4.2 Experimental Results

The results of this experiment generally supported our hypothesis with respect to efficiency. We provide figures that show average values over all users in a particular group, with error bars showing the 95% confidence intervals. The x axis always shows the progression of user's interactions with the system over time: each point is for the $n$th conversation completed by either finding an acceptable restaurant or quitting.

Figure 2 shows that, for the modeling group, the average number of interactions required to find an acceptable restaurant decreased from 8.7 to 5.5, whereas for the control group this quantity actually increased from 7.6 to 10.3. We used linear regression to characterize the trend for each group and compared the resulting lines. The slope for the modeling line differed significantly ($p = 0.017$) from that for the control line, with the former smaller than the latter, as expected.

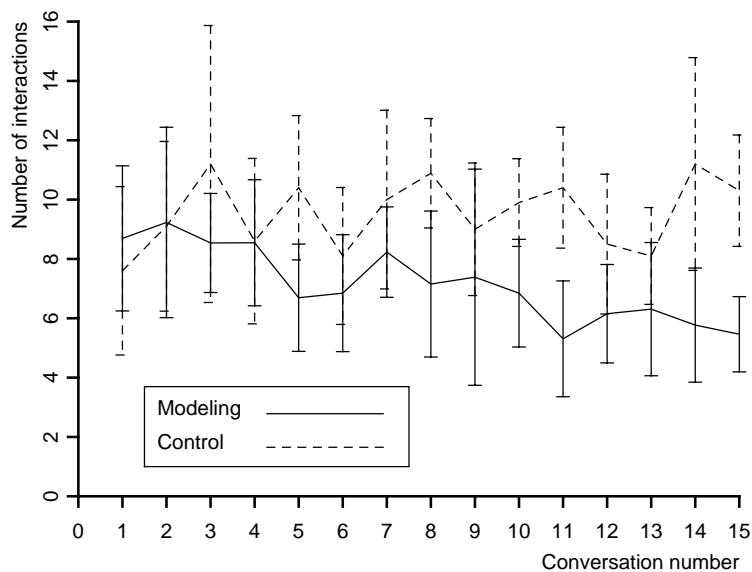

Figure 2: Average number of interactions per conversation.

The difference in interaction times (Figure 3) was even more dramatic. For the modeling group, this quantity started at 181 seconds and ended at 96 seconds, whereas for the control group, it started at 132 seconds and ended at 152 seconds. We again used linear regression to characterize the trends for each group over time and again found a significant difference ($p = 0.011$) between the two curves, with the slope for the modeling subjects being smaller than that for the control subjects. We should also note that these measures include some time for system initialization (which could be up to 10% of the total dialogue time). If we had instead used as the start time the first system utterance of each dialogue, the difference between the two conditions would be even clearer.

The speech recognizer rejected 28 percent of the interactions in our study. Rejections slow down the conversation but do not introduce errors. The misrecognition rate was much lower – it occurred in only seven percent of the interactions in our experiment. We feel both of these rates are acceptable, but expanding the number of supported utterances could reduce the first number further, while potentially increasing the second. In the most common recognition error, the ADAPTIVE PLACE ADVISOR inserted extra constraints that the user did not intend.

The results for effectiveness were more ambiguous. Figure 4 plots the rejection rate as a function of the number of sessions. A decrease in rejection rate over time would mean that, as the system gains experience with the user, it asks about fewer features irrelevant to that user. However, for this dependent variable we found no significant difference ($p = 0.515$) between the regression slopes for the two conditions and, indeed, the rejection rate for neither group appears to decrease with experience. These negative results may be due to the rarity of rejection speech acts in the experiment. Six people never rejected a constraint and on average each person used only 0.53 Reject speech acts after an Attempt-Constrain per conversation (standard deviation = 0.61).

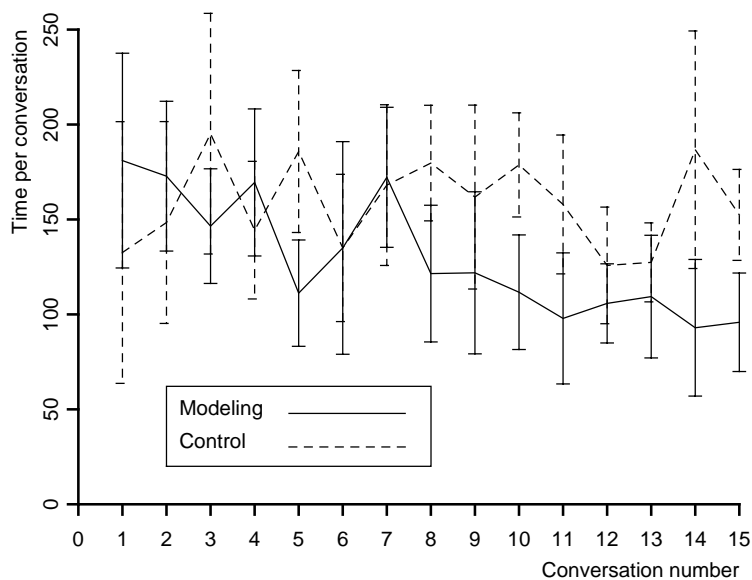

Figure 3: Average time per conversation.

Figure 5 shows the results for hit rate, which indicate that suggestion accuracy stayed stable over time for the modeling group but decreased for the control group. One explanation for the latter, which we did not expect, is that control users became less satisfied with the Place Advisor's suggestions over time and thus carried out more exploration at item presentation time. However, we are more concerned here with the difference between the two groups. Unfortunately, the slopes for the two regression lines were not significantly different ($p = 0.1354$) in this case.

We also analyzed the questionnaire presented to subjects after the experiment. The first six questions (see Appendix A) had check boxes to which we assigned numerical values, none of which revealed a significant difference between the two groups. The second part of the questionnaire contained more open-ended questions about the user's experience with the Adaptive Place Advisor. In general, most subjects in both groups liked the system and said they would use it fairly often if given the opportunity.

### 4.3 Discussion

In summary, our experiment showed that the Adaptive Place Advisor improved the efficiency of conversations with subjects as it gained experience with them over time, and that this improvement was due to the system's update of user models rather than to subjects learning how to interact with the system. This conclusion is due to the significan differences between the user modeling and control groups, for both number of interactions and time per conversation. This significance holds even in the face of large error bars and a small sample size. This in turn implies that the differences are large and the system could make a substantial difference to users.

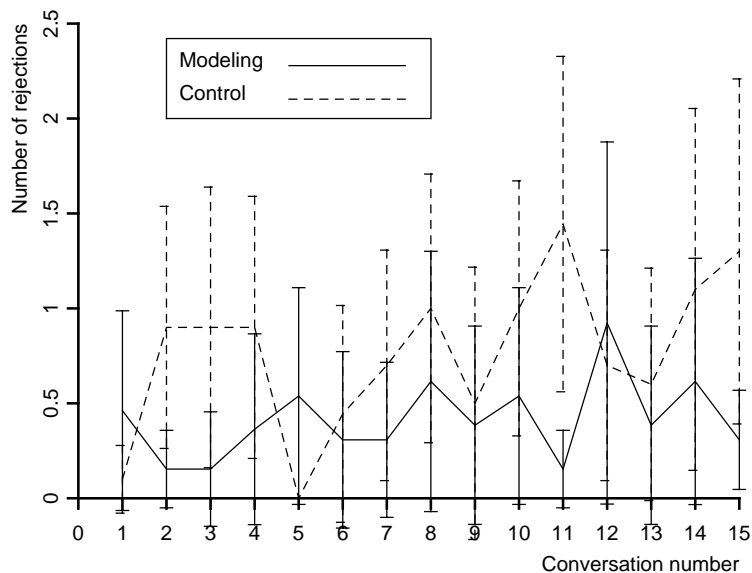

Figure 4: Rejection rate for modeling and control groups.

The results for effectiveness were more ambiguous, with trends in the right direction but no significant differences between the modeling and control groups. Subjects in both conditions generally liked the system, but again we found no significant differences along this dimension. A larger study may be needed to determine whether such differences occur.

Further user studies are warranted to investigate the source of the differences between the two groups. One plausible explanation is that items were presented sooner, on average, in the user modeling group than in the control group. We measured this value (i.e., the average number of interactions before the first item presentation) in the current study and found that it did decrease for the user modeling group (from 4.7 to 3.9) and increased for the control group (from 4.5 to 5.8). This is a reasonably large difference but the difference in slope for the two regression lines is not statistically significant (p=0.165). A larger study may be needed to obtain a significant difference. In general, however, there is an interaction between the user model and the order of questions asked, which in turn influences the number of items matching at each point in the conversation. This in turn determines how soon items are presented in a conversation. Therefore, if items are presented more often in the user modeling group, then the largest influence on the user model is due to item accepts and rejects.

## 5. Related Research

Previous research on related topics can be roughly broken up into three areas, the first focusing on personalized recommendation systems, the second on conversational interfaces, and the third on adaptive dialogue systems. We restrict our discussion here to the most strongly related work.

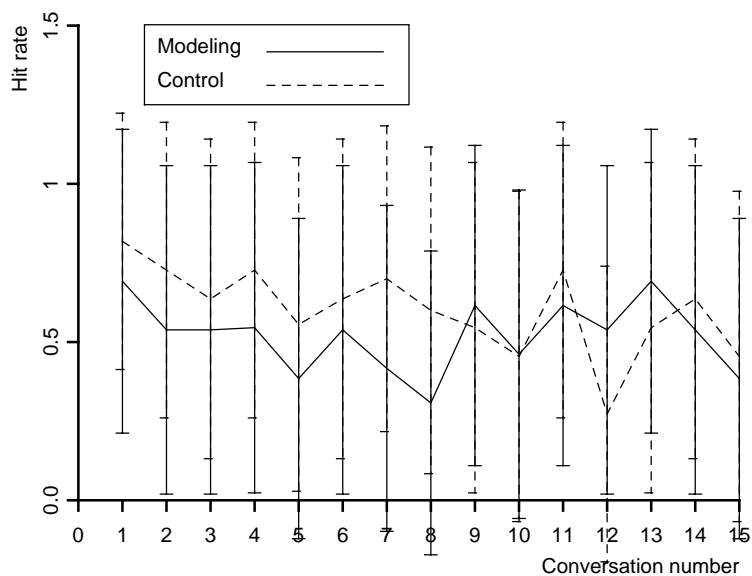

Figure 5: Hit rate for modeling and control groups.

## 5.1 Personalized Recommendation Systems

Although research in personalized recommendation systems has become widespread only in recent years, the basic idea can be traced back to Rich (1979), discussed below with other work on conversational interfaces. Langley (1999) gives a more thorough review of recent research on the topic of adaptive interfaces and personalization.

Several other adaptive interfaces attempt to collect user information unobtrusively. An interesting example is the CASPER project (Rafter et al., 2000), an online recruitment service. The project investigates methods that translate click and read-time data into accurate relevancy information, given that the raw data is inherently noisy. Similarly, Goecks and Shavlik (2000) describe a technique for learning web preferences by observing a user's browsing behavior. Another example is the ADAPTIVE ROUTE ADVISOR (Rogers et al., 1999), which recommends driving routes to a specified destination. The system collects preferences about attributes such as number of turns and driving time on the basis of the user's selections and modifications of the system's proposed routes.

While the ADAPTIVE PLACE ADVISOR uses constraint-based interaction to search for items, this is not the only interaction approach for item search. An alternative is taken by the *candidate/critique*, or *tweaking*, approach. Tweaking systems, such as the FIND ME suite (Burke, 1999), typically require the user to begin an interaction by filling in values for a few predetermined attributes. They then present an item, at which point the user has the opportunity to change some search parameters to try to find a more desirable item. Eaton, Freuder, and Wallace (1997) take a similar approach in their MATCHMAKING system. In addition, they exploit constraint satisfaction to manage the search. Neither the FIND ME suite nor the MATCHMAKING system, however, learns user models. A related system that does include a learning component is that of Shearin and Lieberman (2001), which learns attribute preferences unobtrusively. While tweaking is a valid method, it is not appropriate, we feel, in an environment in which speech is the only interaction mode, since presenting the user with his options would be somewhat cumbersome. Even though our current system also presents options once the search is constrained, it limits the number of items presented. Even in a "full speech" version this does not seem onerous.

## 5.2 Conversational Interfaces

There is considerable ongoing work in the area of conversational systems, as evidenced in the general surveys by Dybkjær et al. (2000) and Maier et al. (1996). Zukerman and Litman (2001) give a more thorough overview of user modeling in dialogue systems. Rich (1979) reported one of the earliest (typewritten) conversational interfaces, which focused on book recommendation. At the beginning of an interaction, the system asked several questions to place the user in a stereotype group, thereby initializing the user model. As each conversation progressed, this model was adjusted, with the system using ratings to represent uncertainty. However, the language understanding capabilities of the system were limited, mostly allowing only yes/no user answers. More recently, dialogue systems utilize models of user's beliefs and intentions to aid in dialogue management and understanding, though typically these systems maintain models only over the course of a single conversation (Kobsa & Wahlster, 1989).

As noted in Section 2.3, an important distinction is whether only one conversational participant keeps the initiative, or whether the initiative can switch between participants. Two ambitious mixed-initiative systems for planning tasks are TRAINS (Allen et al., 1995) and more recent TRIPS (Allen et al., 2001). Like the PLACE ADVISOR, these programs interact with the user to progressively construct a solution, though the knowledge structures are partial plans rather than constraints, and the search involves operators for plan modification rather than for database contraction and expansion. TRAINS and TRIPS lack any mechanism for user modeling, but the underlying systems are considerably more mature and have been evaluated extensively.

Smith and Hipp (1994) describe another related mixed-initiative system with limited user modeling, in this case a conversational interface for circuit diagnosis. Their system aims to construct not a plan or a set of constraints, but rather a proof tree. The central speech act, which requests knowledge from the user that would aid the proof process, is invoked when the program detects a 'missing axiom' that it needs for its reasoning. This heuristic plays the same role in their system as does the PLACE ADVISOR's heuristic for selecting attributes to constrain during item selection. The interface infers user knowledge during the course of only a single conversation, not over the long term as in our approach.

With respect to dialogue management, several previous systems have used a method similar to our frame-based search. In particular, Seneff et al. (1998) and Dowding et al. (1993) developed conversational interfaces that give advice about air travel. Like the PLACE ADVISOR, their systems ask the user questions to reduce the number of candidates, treating flight selection as the interactive construction of database queries. However, the question sequence is typically fixed in advance, despite the clear differences among individuals in this domain. Also, these systems usually require that all constraints be specified before item presentation begins.

An alternative technique for selecting which questions to ask during information elicitation is presented in Raskutti and Zukerman (1997). Their overall system necessitates that the system recognize plans the user is attempting to carry out. Then the system must decide how to best complete those plans. When insufficient information is available for plan formation, their system enters an information seeking subdialogue similar to the constraint-satisfaction portion of our dialogues. Their system can decide which question to ask based on domain knowledge or based on the potential informativeness of the question.

Another approach to dialogue management is "conversational case-based reasoning" (Aha, Breslow & Muñoz-Avila, 2001), which relies on interactions with the user to retrieve cases (items) that will recommend actions to correct some problem. The speech acts and basic flow of control have much in common with the ADAPTIVE PLACE ADVISOR, in that the process of answering questions increasingly constrains available answers. One significant difference is that their approach generates several questions or items, respectively, at a time, and the user selects which question to answer or which item is closest to his or her needs, respectively.

Finally, our approach draws on an alternative analysis of item recommendation, described by Elio and Haddadi (1998, 1999). The main distinctions from that work are that their approach does not include personalization, that they distinguish between search through a task space and through a discourse space, while we combine the two, and that they place a greater emphasis on user intentions. Keeping a distinction between the task and the discourse space in a personalized system would unnecessarily complicate decisions about when to perform user model updates and about how to utilize the model.

## 5.3 Adaptive Dialogue Systems

Finally, another body of recent work describes the use of machine learning or other forms of adaptation to improve dialogue systems.[10] Researchers in this area develop systems that learn user preferences, improve task completion, or adapt dialogue strategies to an individual during a conversation.

The closest such work also pursues our goal of learning user preferences. Carberry et al. (1999) report one such example for consultation dialogues, but take a different approach. Their system acquires value preferences by analyzing both user's explicit statements of preferences and their acceptance or rejection of the system's proposals. It uses discrete preference values instead of our more fine-grained probability model. Also, their system does not use preferences during item search but only at item presentation time to help evaluate whether better alternatives exist. Finally, their evaluation is based on subject's judgements of the quality of the system's hypotheses and recommendations, not on characteristics of actual user interactions. We could, however, incorporate some of their item search ideas, allowing near misses between user-specified constraints and actual items.

Another system that focuses on user preferences is an interactive travel assistant (Linden et al., 1997) that carries out conversations via a graphical interface. The system asks questions with the goal of narrowing down the available candidates, using speech acts similar to ours, and also aims to satisfy the user with as few interactions as possible. Their approach to minimizing the number of interactions is to use a candidate/critique approach. From a user's responses, the system infers a model represented as weights on attributes such as price and travel time. Unlike the ADAPTIVE PLACE ADVISOR, it does not carry these profiles over to future conversations, but one can envision a version that does so.

---

10. The work on adaptation of speech recognition grammars (e.g., Stolcke et al., 2000), while related, addresses a different problem and uses different learning techniques, so we do not discuss it here.

Several authors use reinforcement learning techniques to improve the probability of or process of task completion in a conversation. For example, Singh et al. (2002) use this approach to determine the system's level of initiative and the amount of confirmation of user utterances. Their goal is to optimize, over all users, the percentage of dialogues for which a given task is successfully completed. This system leverages the learned information when interacting with all users, rather than personalizing the information. Also, Levin, Pieraccini, and Eckert (2000) use reinforcement learning to determine which question to ask at each point during an information seeking search, but do not demonstrate the utility of their approach with real users.

Finally, a number of systems adapt their dialogue management strategy over the course of a conversation based on user responses or other dialogue characteristics. For example, Litman and Pan (2002) use a set of learned rules to decide whether a user is having difficulty achieving their task, and modify the level of system initiative and confirmation accordingly. Maloor and Chai (2000) present a help-desk application that first classifies the user as a novice, moderate, or expert based on responses to prompts. It then adjusts the complexity of system utterances, the jargon, and the complexity of the path taken to achieve goals. Horvitz and Paek (2001) apply user modeling to a dialogue system that uses evidence from the current context and conversation to update a Bayesian network. The network influences the spoken language recognition hypothesis and causes appropriate adjustments in the system's level of initiative. Chu-Carroll (2000) describes a system that adapts both language generation and initiative strategies for an individual user within a single dialogue. Also, Jameson et al. (1994) use Bayesian networks in a system that can take the role of either the buyer or seller in a transaction, and that changes its inquiry or sales strategy based on beliefs inferred from the other participant's utterances.

## 6. Directions for Future Work

Our results to date with the ADAPTIVE PLACE ADVISOR are promising but much remains to be done. In this section, we discuss ways to make the search model more flexible, to expand the conversational model, and to enrich the user model and learning technique. We also consider more extensive evaluations of the system.

### 6.1 Search Model

With respect to the search mechanism, we first plan to investigate alternative techniques for using item similarity values to determine which to return, for example by cutting off items at the point at which similarity drops off most steeply, instead of our current use of a threshold. We also note that work such as that of Cohen, Schapire, and Singer (1999) on learning to rank instances could apply nicely to this work, augmenting our current item ranking scheme. Additionally, we plan develop a version of the system that generates alternative items or values in an over-constrained situation (Qu & Beale, 1999). One way to do this would be to use the preferences to estimate the strength of a stated constraint, or to merge our preference-based similarity metric with a more traditional domain-specific similarity metric (Pieraccini et al., 1997). We also plan to evaluate the effect of making even stronger assumptions about user preferences. For example, if the system is certain enough about a value preference, it may not have to ask a question about the associated attribute.

A final improvement of the search mechanism concerns the techniques for ranking attributes for constraining and relaxing. For attribute constraint ranking, we have implemented but not yet evaluated a conditional entropy measure (Göker & Thompson, 2000). The system selects the attribute to constrain by determining the attribute with the highest conditional entropy among the unconstrained attributes. This scheme would not be useful for ranking attributes to relax. Therefore, the system simply determines the size of the case base that would result if each attribute were relaxed, ranks these case bases from smallest to largest, and orders the attributes accordingly, excluding those attributes that, if relaxed, would still result in an empty case base. We also plan to investigate the combination of the user model with information gain, as well as with alternative attribute ranking techniques such as the one used by Abella, Brown, and Buntschuh (1996). Another option is to add personalization to or otherwise adapt the variable selection techniques used by constraint-satisfaction solvers.

## 6.2 Conversational Model

We plan to progress towards more complex dialogues with more complex constraints. First, we plan to increase the number of speech acts available to the user. For example, we will add confirmation dialogues and improve the current clarification dialogues, thus allowing other types of adaptation strategies, as in Singh et al. (2002). In a longer term investigation, we plan to extend our adaptation techniques to handle more complex travel planning dialogues (Seneff, Lau, & Polifroni, 1999; Walker & Hirschman, 2000). These may require additions to the user model, such as preferences regarding language and dialogue style, including initiative, system verbosity, and vocabulary. These will in turn need to be appropriately acquired and utilized by the system. In general, the insights we have already gained into utilizing and acquiring user preferences at different junctures of the dialogue and search process should prove useful in supporting personalization in other tasks.

## 6.3 User Model

To improve the user model, we first plan to add more types of preferences. As we discussed in Section 3.1, combination and diversity preferences can capture more complex user behavior than our current model, and we plan to incorporate both into the next version of our system. Combination preferences can help it better predict either values or acceptable attributes, based on previously provided constraints. The PLACE ADVISOR can model value combination preferences by learning association rules (Agrawal, Imielinski, & Swami, 1993) or extending to a Bayesian network, either of which would then influence the query, in turn influencing the similarity calculation and the case base. For preferences about acceptable attribute combinations, the system can learn conditional probabilities based on past interactions and use this to influence attribute ranking.

While "drifting" preferences are not likely to cause problems in item selection applications as they might in ones like news updates, our model could be extended to handle within-user diversity. One way to do this is to capture the user's desired time interval between the suggestion of a particular item or value. We can calculate this by determining the mean time interval between a user's explicit selection or rejection of a value (value diversity preferences) or item (item diversity preferences). We will incorporate both these diversity preferences into the similarity calculation from Section 3.2 by extending $R_I$ and $P(V_j)$ in that equation to incorporate time effects. We define $R_D(I)$ and $P_D(V_j)$ as:

$$R_D(I) = R_I \times \frac{1}{1 + e^{-k_I(t - t_I - t_{ID})}}$$

$$P_D(V_j) = P(V_j) \times \frac{1}{1 + e^{-k_V(t - t_V - t_{VD})}} \ ,$$

where $t$ is the current time, $t_I$ and $t_V$ are the time when the item or value was last selected, and $t_{ID}$ and $t_{VD}$ are the time differences the user wants to have between having the item or value suggested again. $R_D$ and $P_D$ are in form of a sigmoid function where $k_I$ and $k_V$ determine the curve's slope. One empirical question is whether users also have attribute diversity preferences. We hypothesize that diversity preferences differ for each value of each attribute, and that this implicitly overrides attribute diversity. For example, a user may have different preferences about the frequency with which expensive restaurants versus cheap ones are suggested, but may not care about how often questions about price are asked. We plan to investigate this hypothesis.

There are other improvements we might add to our user modeling technique. For example, the system may learn more quickly if it updates the user model in dialogue situations other than the current three. Also, using collaborative user models to initialize individual models could speed up the learning process. A more explicit combination of collaborative and individual user models (Melville, Mooney, & Nagarajan, 2002; Jameson & Wittig, 2001) is also a viable direction to explore.

## 6.4 Evaluation

Finally, we are planning to carry out a larger user study. We must further verify that the differences in our study were not due to task difficulty differences since we did not control for the difficulty of finding a particular item. In particular, different values for the same constraints may not result in the same number of matching items. So even though two users have answered the same number of questions, the number of matching items for one user may be small enough for the system to begin presenting them, while the other user may need to answer additional questions first. To support an expanded evaluation, we have implemented a version of the system that recommends movies, which will let us draw from a broader user base. This should help us measure user satisfaction more easily, as Walker et al. (1998) have noted that efficiency is not the only important consideration, and that users might tend to prefer more predictable interfaces.

## 7. Conclusions

In this paper, we described an intelligent adaptive conversational assistant designed to help people select an item. Overall, we made significant inroads into methods for unobtrusively acquiring an individual, long term user model during recommendation conversations. We expanded on previous work on adaptive recommendation systems that were not conversational, and on dialogue systems that were not user adaptive. Our long-term goal is to develop even more powerful methods, capable of adapting to a user's needs, goals, and preferences over multiple conversations. While we leveraged off the feedback between conversation and recommendation, such feedback is likely to be present in other tasks such as planning or scheduling.

The two key problems addressed by our research are the design of adaptive recommendation systems when conversations are the interaction mode, and the addition of personalization to dialogue systems, starting here with dialogues for recommendation. Thus, unlike many recommendation systems that accept keywords and produce a ranked list, this one carries out a conversation with the user to progressively narrow his options. In solving these problems, we introduced a novel approach to the acquisition, use, and representation of user models. Unlike many other adaptive interfaces, our system constructs and utilizes user models that include information beyond complete item preferences. This is key for the support of personalization in conversations. We used a relatively simple model of dialogue to focus on the issues involved in personalization. We also described experimental results showing the promise of our technique, demonstrating a reduction in both the number of interactions and in the conversation time for users interacting with our adaptive system when compared to a control group.

Of course, there are still several open questions and opportunities for improvement. The user model, conversational model, and search models are functional but we plan to improve them further. We are also extending our conversational approach to items other than destinations, such as books and movies, and we plan to link the system to other assistants like the Adaptive Route Advisor (Rogers et al., 1999). Our goal for such additions is to provide new functionality that will make the ADAPTIVE PLACE ADVISOR more attractive to users, but also to test the generality of our approach for adaptive recommendation. In turn, this should bring us closer to truly flexible computational aides that carry out natural dialogues with humans.

## Acknowledgments

This research was carried out while the first author was at the Center for the Study of Language and Information, Stanford University, and the other authors were at the DaimlerChrysler Research and Technology Center in Palo Alto, California. We thank Renée Elio, Afsaneh Haddadi, and Jeff Shrager for the initial conception and design of the ADAPTIVE PLACE ADVISOR, Cynthia Kuo and Zhao-Ping Tang for help with the implementation effort, and Stanley Peters for enlightening discussions about the design of conversational interfaces. Robert Mertens and Dana Dahlstrom were crucial in carrying out the user studies.

## Appendix A. Questionnaire

1. What did you think about the interaction with the system, did it

```
0    1    2    3    4    5    6    7    8
talk               right               not
too                amount of           enough
much               talking             talking
```

2. How easy was it to find a restaurant you liked?

```
0    1    2    4
very          not
easy          easy
```

3. Did the system deliver restaurants you liked?

```
 0    1    2    4
yes             no
```

4. Please rate the interaction with the system on a scale between standard human-computer-interaction and a person to person conversation via telephone.

```
0    1    2    3    4    5    6
human-                   phone
computer                 conversation
interaction
```

5. Do you think the APA is a useful system?

```
 0    1    2    3    4
yes             no
```

6. Do you think the conversation was significantly more distracting than a similar conversation with a real person?

```
 0    1    2    3    4
no              yes
```